\documentclass{appolb}
\usepackage{graphicx}

\begin{document}
\title{
A short review of some double-parton scattering processes%
\thanks{Presented at Excited QCD 2015, Tatranska Lomnica, Slovakia, 8-14
March 2015}%
}
\author{Antoni Szczurek
\address{Institute of Nuclear Physics (PAN), Krak\'ow, Poland,\\
Rzesz\'ow University, Rzesz\'ow, Poland}
\\
}

\maketitle

\begin{abstract}
A few examples of quickly developing field of double
parton scattering are discussed.
We present arguments that the production of two pairs of charm
quark-antiquark is the golden reaction to study the double parton
scattering effects.  
We discuss briefly also mechanism of single parton
scattering and show that it gives much smaller contribution
to the $c \bar c c \bar c$ final state.
In this context we discuss also the contribution of perturbative
parton-splitting mechanism which 
should also be added to the conventional DPS mechanism. 
The presence of the letter leads to collision energy and other
kinematical variables dependence of so-called $\sigma_{eff}$.
We briefly discuss production of four jets. We concentrate on estimation
of the contribution of DPS for jets widely separated in rapidity.
Finally we briefly mention about DPS effects in production of $W^+ W^-$.
\end{abstract}

\section{Introduction}

The double-parton scattering was recognized already 
in seventies and eighties 
\cite{LP78,T1979,GSH80,H1983,PT1984,PT1985,M1985,HO1985,SZ1987}.  
Several estimates of the cross section for different processes have been
presented in recent years \cite{DH1996,KS2000,FT2002,BJS2009,GKKS2010,
SV2011,BDFS2011,KKS2011,BSZ2011}. The theory of the double-parton 
scattering is quickly developing
\cite{S2003,KS2004,SS2004,GS2010,GS2011,DS2011,RS2011,DOS2011,MW2012a,
MW2012b,DK2013,DKK2014}.

It was recognized recently that the production of $c \bar c c \bar c$
is a very good place to study DPS effects \cite{LMS2012}.
Here, the quark mass is small enough to assure that the cross section 
for DPS is large, but large enough that each of the scatterings 
can be treated within pQCD.
In the meantime the LHCb collaboration presented interesting 
data for simultaneous production of two charmed mesons 
\cite{Aaij:2012dz}.  
In Ref.~\cite{MS2013} we discussed that the large cross section
in \cite{Aaij:2012dz} is a 
footprint of double parton scattering.

25 years ago Mueller and Navelet predicted strong decorrelation 
in relative azimuthal angle \cite{Mueller:1986ey} of jets with large
rapidity separation due to exchange of the BFKL ladder between quarks. 
Since then both leading-logarithmic
\cite{Mueller:1986ey,DelDuca:1993mn,Stirling:1994he,DelDuca:1994ng,Kim96,Andersen2001}
and higher-order BFKL effects
\cite{Bartels-MNjets,Vera:2007kn,Marquet:2007xx,Colferai:2010wu,Caporale:2011cc,Ivanov:2012ms,
Caporale:2012ih,Ducloue:2013hia,Ducloue:2013bva,DelDuca2014} 
were calculated and discussed.
The effect of the NLL correction is large and leads to significant
lowering of the cross section.
The LHC opens a new possibility to study the decorrelation in azimuthal angle. 
First experimental data measured at $\sqrt{s}$ = 7 TeV are expected 
soon \cite{CMS_private}.
We discussed recently the contribution of DPS to the jets
widely separated in rapidity \cite{MS2014_DPSjets}.

The double parton scattering mechanism of $W^+ W^-$ production
was discussed e.g. in 
Refs.~\cite{KS2000,Kulesza2010,LSR2015,GL2014}. 
The $W^+ W^-$ final states constitutes a background to Higgs production.
It was discussed recently that double-parton scattering could explain
a large part of the observed signal \cite{KP2013}. Here we discuss
briefly the double parton scattering mechanism of $W^+W^-$ production. 

\section{Sketch of the formalism}

\begin{figure}
\begin{center}
\includegraphics[width=4cm]{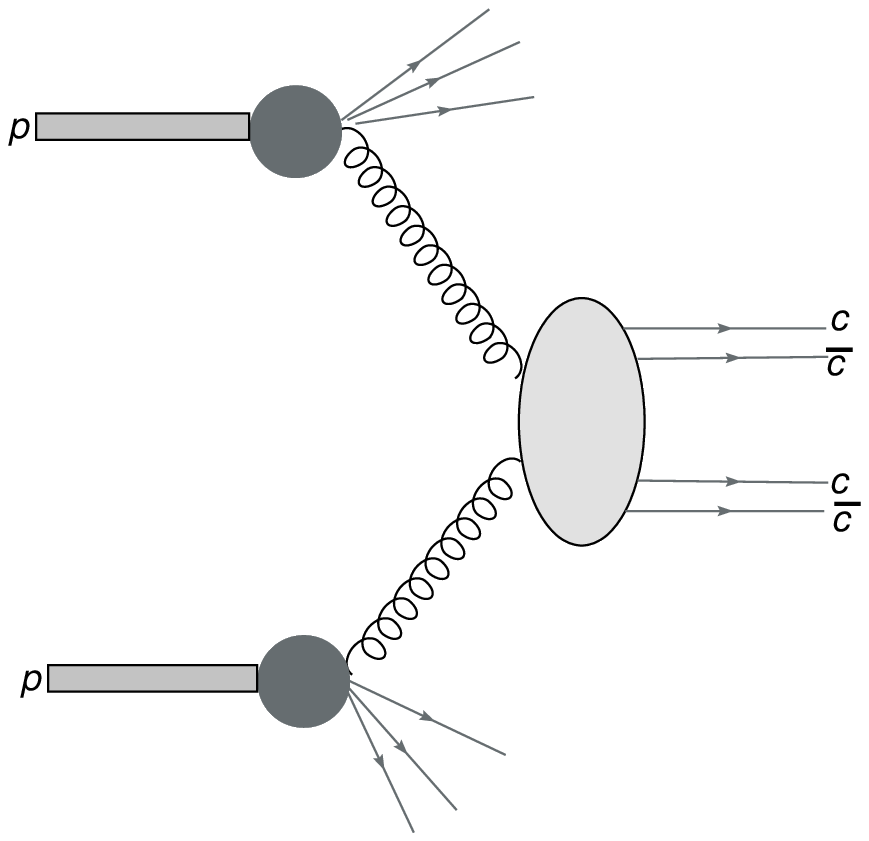}
\includegraphics[width=5cm]{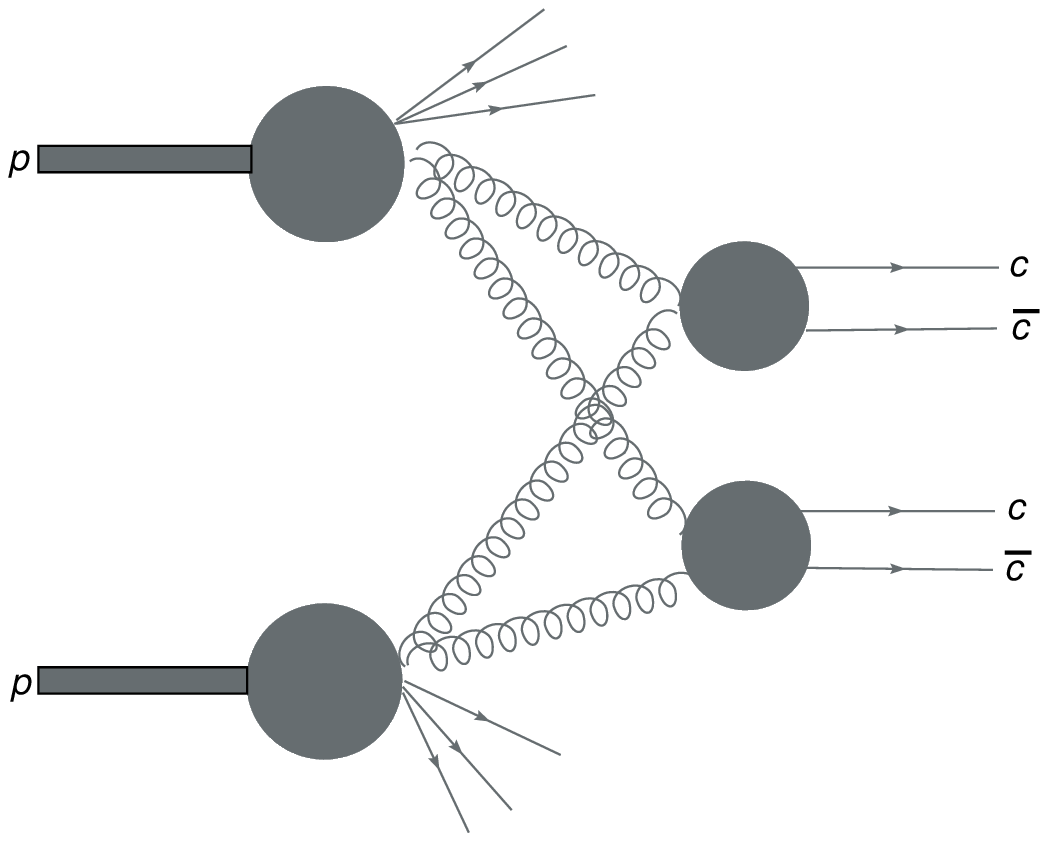}
\end{center}
\vskip -0.3cm
\caption{
SPS and DPS production mechanisms of $c \bar c c \bar c$. 
}
\label{fig:diagrams:ccbarccbar}
\end{figure}

\begin{figure}[!h]
\begin{center}
\includegraphics[width=4.0cm]{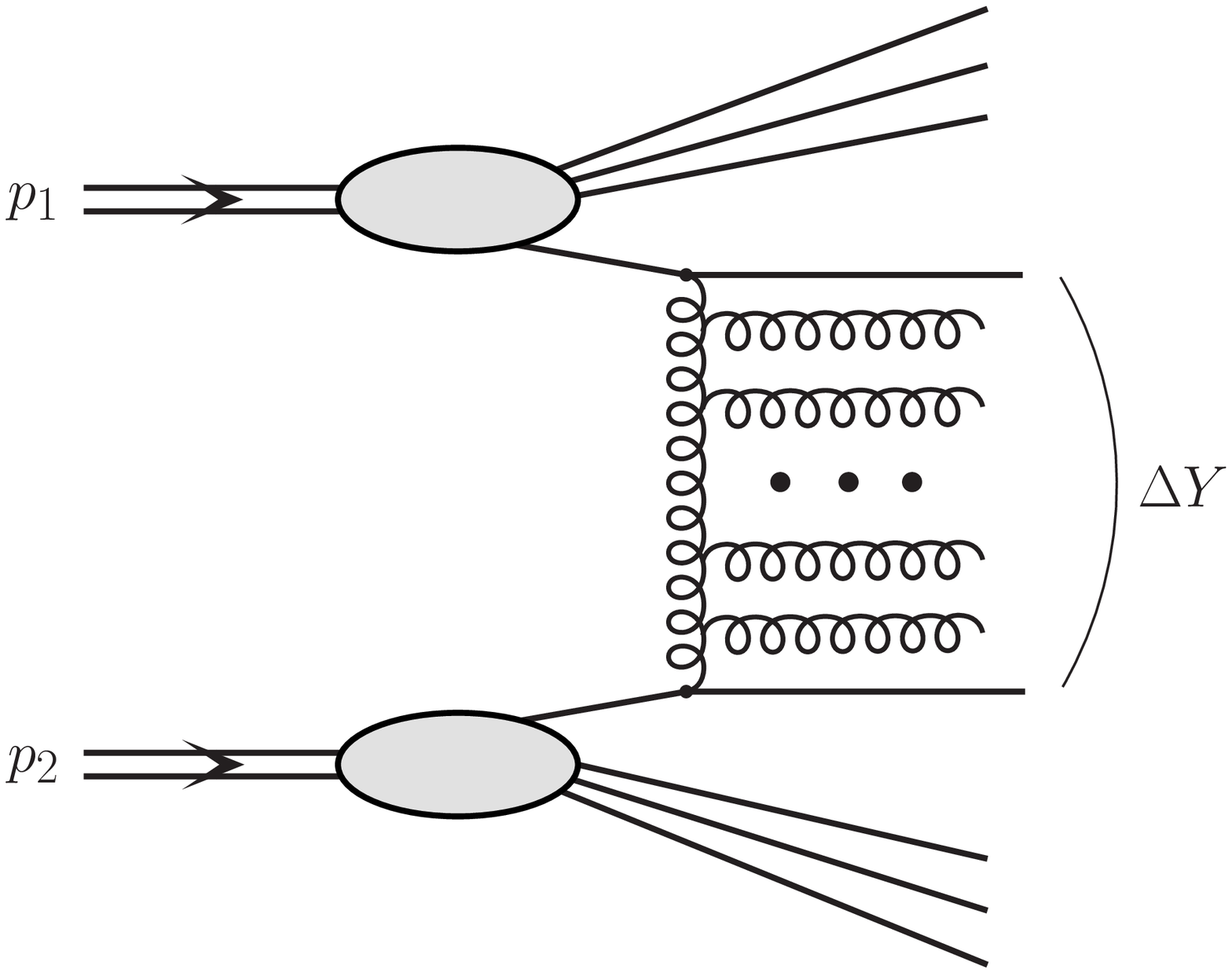}
\includegraphics[width=4.0cm]{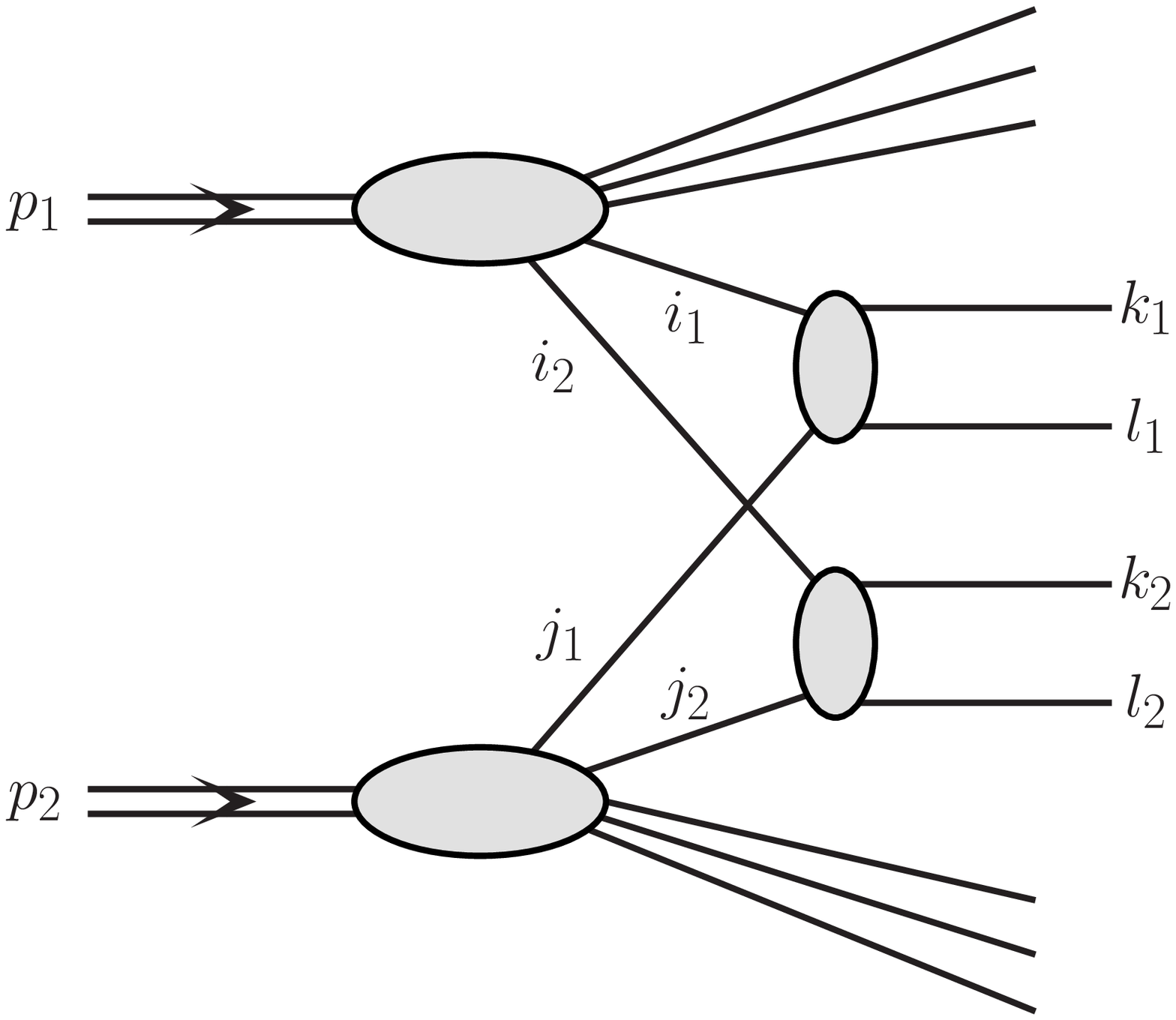}
\end{center}
\vskip -0.3cm
\caption{
\small A diagramatic representation of the Mueller-Navelet jet
production (left diagram) and of the double paron scattering mechanism
(right diagram).
}
 \label{fig:diagrams}
\end{figure}

Let us consider for example production of $c \bar c c \bar c$ final state
within the DPS framework. In a simple probabilistic picture the 
cross section for double-parton scattering can be written as
(see \cite{MS2013}):
\begin{equation}
\sigma^{DPS}(p p \to c \bar c c \bar c X) = \frac{1}{2 \sigma_{eff}}
\sigma^{SPS}(p p \to c \bar c X_1) \cdot \sigma^{SPS}(p p \to c \bar c X_2).
\label{basic_formula}
\end{equation}
The simple formula (\ref{basic_formula}) can be generalized to address 
differential distributions. 
A more general formula for the cross section can be written formally 
in terms of double-parton distributions, e.g. $F_{gg}$, $F_{qq}$, etc.
(see e.g.\cite{szczurek_epiphany2015}).
 
\begin{figure}[!h]
\begin{center}
\includegraphics[width=4cm]{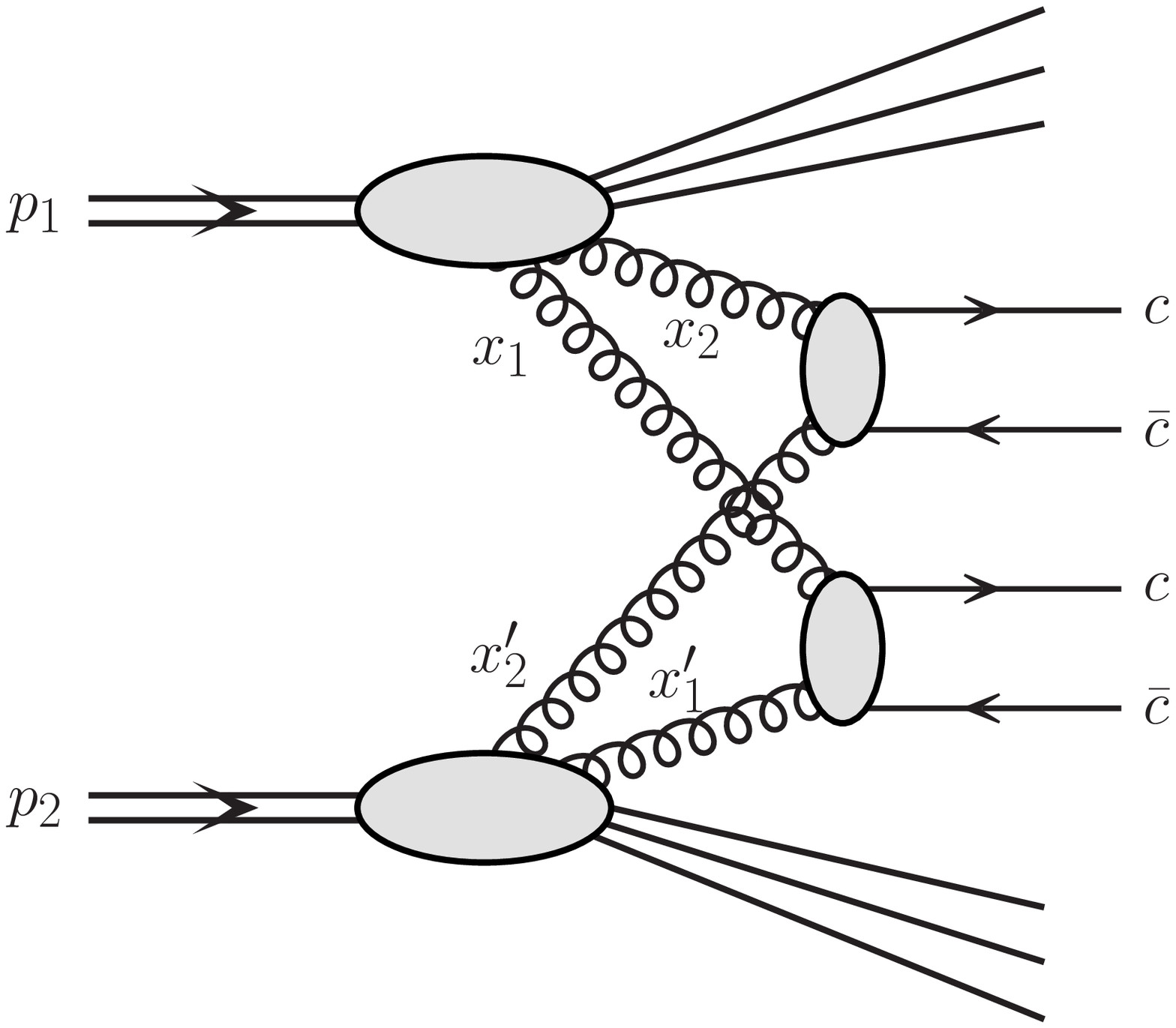}
\includegraphics[width=4cm]{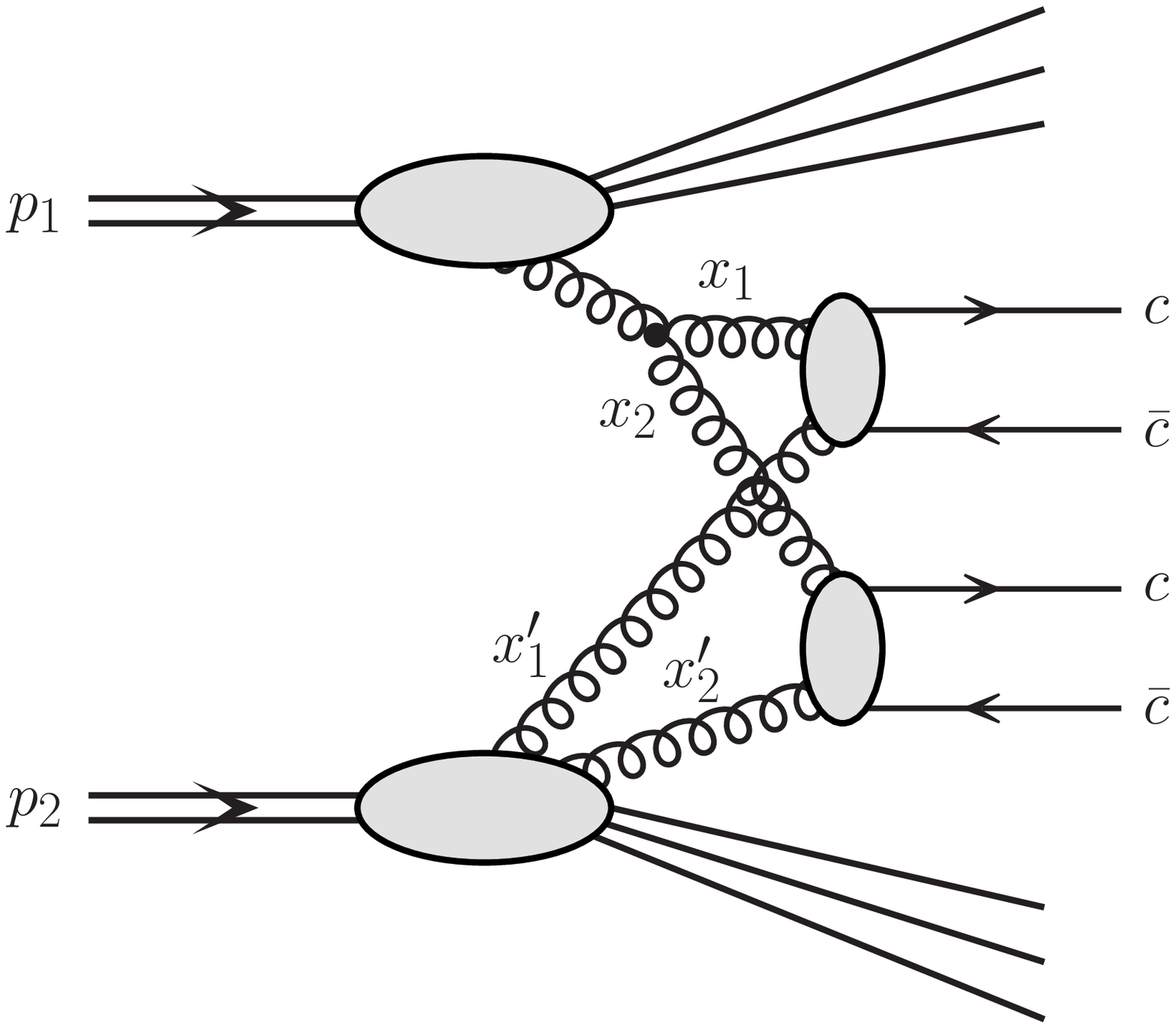}
\includegraphics[width=4cm]{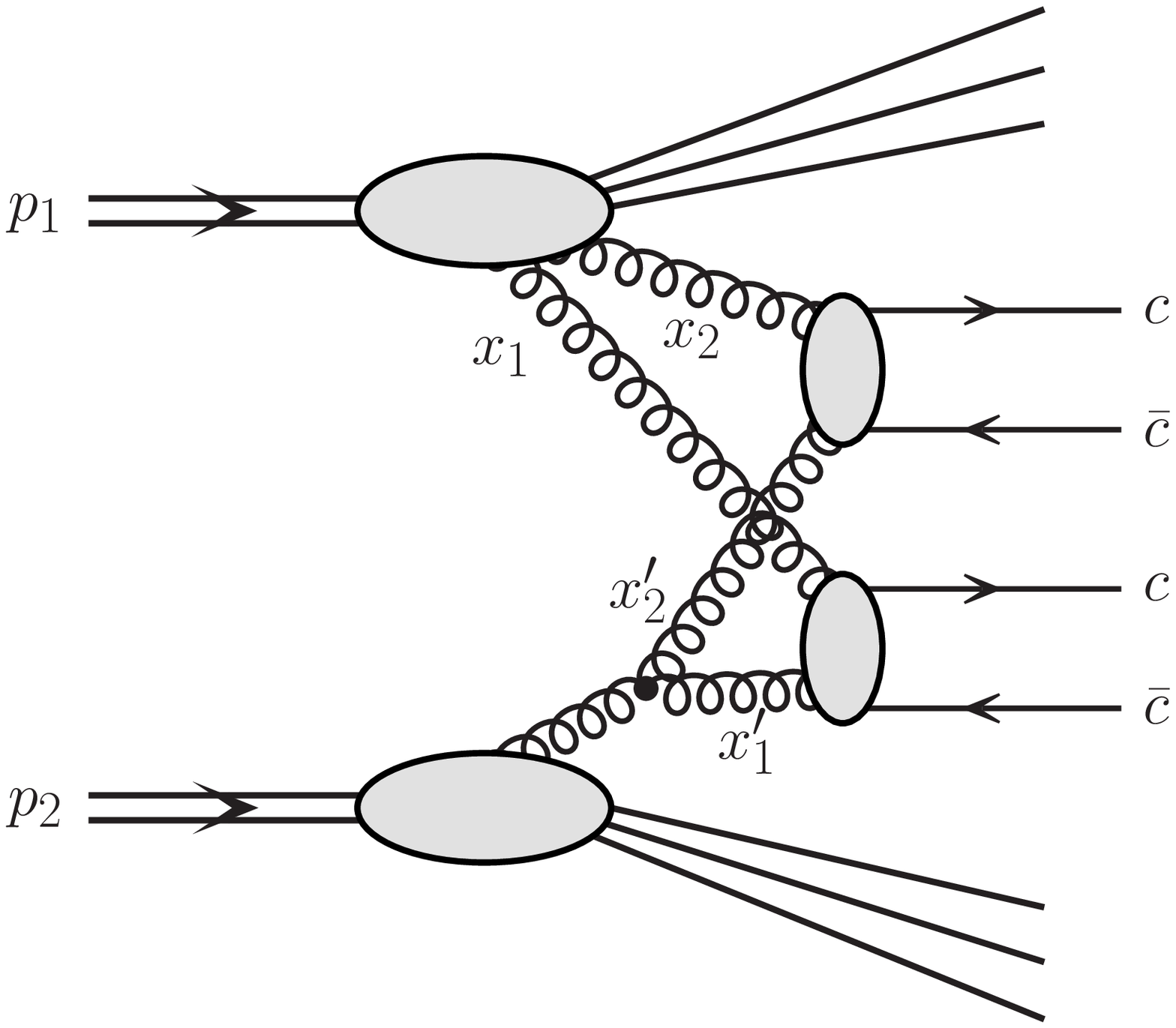}
\end{center}
\vskip -0.3cm
   \caption{
\small The diagrams for DPS production of $c \bar c c \bar c$.
}
 \label{fig:diagrams_ccbarccbar}
\end{figure}

In the $k_t$-factorization approach the differential cross section for
DPS production of $c \bar c c \bar c$ system can be written as: 
\begin{eqnarray}
\frac{d \sigma^{DPS}(p p \to c \bar c c \bar c X)}{d y_1 d y_2 d^2 p_{1,t} d^2 p_{2,t} 
d y_3 d y_4 d^2 p_{3,t} d^2 p_{4,t}} = \nonumber \;\;\;\;\;\;\;\;\;\;\;\;\;\;\;\;\;\;\;\;\;\;\;\;\;\;\;\;\;\;\;\;\;\;\;\;\;\;\;\;\;\;\;\;\\ 
\frac{1}{2 \sigma_{eff}} \cdot
\frac{d \sigma^{SPS}(p p \to c \bar c X_1)}{d y_1 d y_2 d^2 p_{1,t} d^2 p_{2,t}}
\cdot
\frac{d \sigma^{SPS}(p p \to c \bar c X_2)}{d y_3 d y_4 d^2 p_{3,t} d^2 p_{4,t}}.
\end{eqnarray}
In Fig.~\ref{fig:diagrams_ccbarccbar} we illustrate a conventional
and perturbative splitting DPS mechanisms
for $c \bar c c \bar c$ production. The 2v1 single parton splitting mechanism 
(the second and third diagrams in the figure) were considered first 
in \cite{GMS2014}.

Some more details of the formalism are given e.g.
in \cite{szczurek_epiphany2015}.

\section{Examples of results}

\begin{figure}[!h]
\begin{center}
\includegraphics[width=5.0cm]{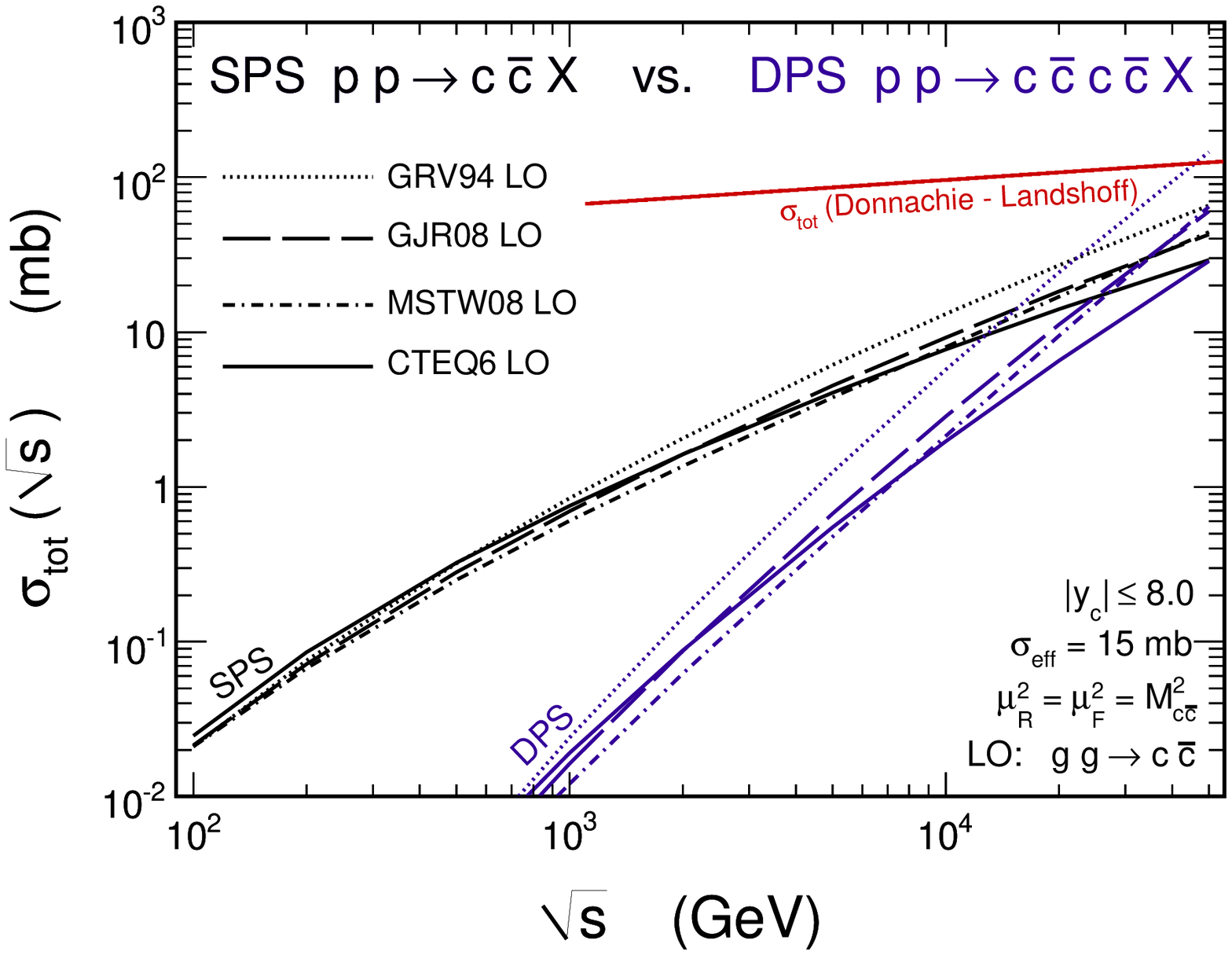}
\includegraphics[width=5.0cm]{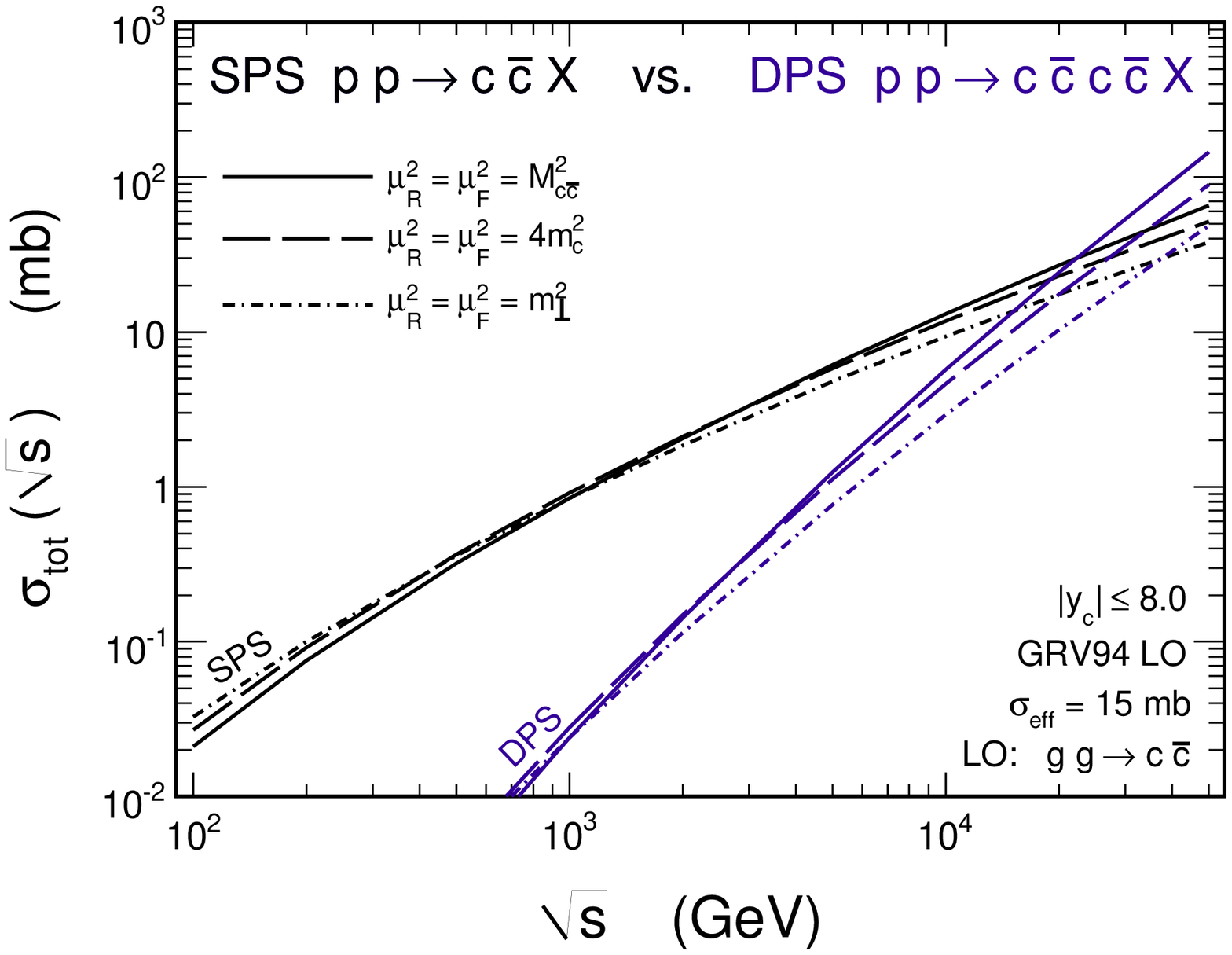}
\end{center}
\vskip -0.3cm
\caption{
\small Total LO cross section for $c \bar c$ and double-parton
scattering production of $c \bar c c \bar c$ as a function of
center-of-mass energy (left panel) and uncertainties due to 
the choice of (factorization, renormalization) scales (right panel). 
We show in addition a parametrization of the total cross section 
in the left panel.
}
 \label{fig:single_vs_double_LO}
\end{figure}

In Fig.~\ref{fig:single_vs_double_LO} we
compare cross sections for the single and double-parton
scattering as a function of proton-proton center-of-mass energy. 
At low energies the single-parton scattering dominates. For reference 
we show the proton-proton total cross section as a function
of collision energy as parametrized in Ref.~\cite{DL92}.
At low energy the $c \bar c$ or $ c \bar c c \bar c$ cross sections are much
smaller than the total cross section. At higher energies both 
the contributions approach the total cross section.
At LHC energies the cross section for both terms become comparable.
This is a completely new situation.

\begin{figure}[!h]
\begin{center}
\includegraphics[width=6cm]{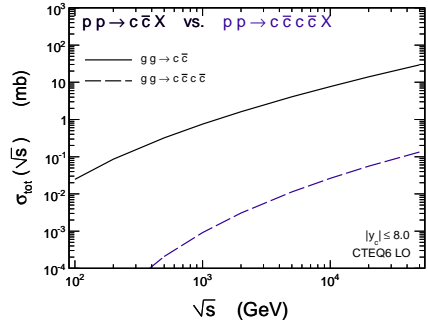}
\end{center}
\vskip -0.3cm
\caption{Cross section for SPS production of $c \bar c c \bar c$
compared to this for standard $c \bar c$ production as a function
of collision energy.
}
\label{fig:SPSccbar_vs_SPSccbarccbar}
\end{figure}

So far we have concentrated on DPS production of $c \bar c c \bar c$
and completely ignored SPS production of $c \bar c c \bar c$. In
Refs.\cite{SS2012,Hameren2014} we calculated the SPS contribution
in high-energy approximation \cite{SS2012} and including all diagrams
in the collinear-factorization approach \cite{Hameren2014}.
In Fig.~\ref{fig:SPSccbar_vs_SPSccbarccbar} we show the cross section
from Ref.~\cite{Hameren2014}. The corresponding cross section at 
the LHC energies is more than two orders of magnitude smaller than 
that for $c \bar c$ production and much smaller than the DPS
contribution.

\begin{figure}
\begin{center}
\includegraphics[width=5.5cm]{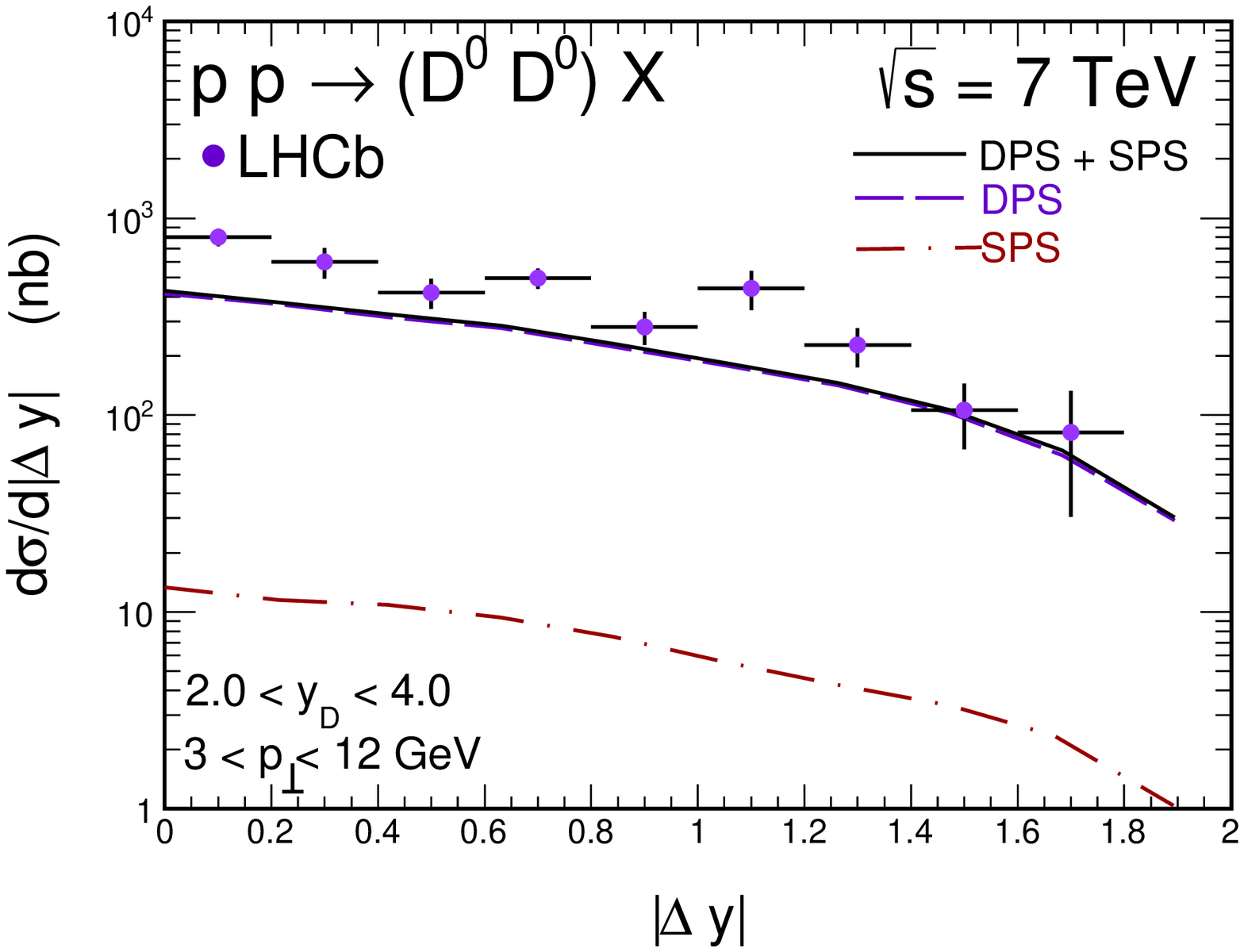}
\includegraphics[width=5.5cm]{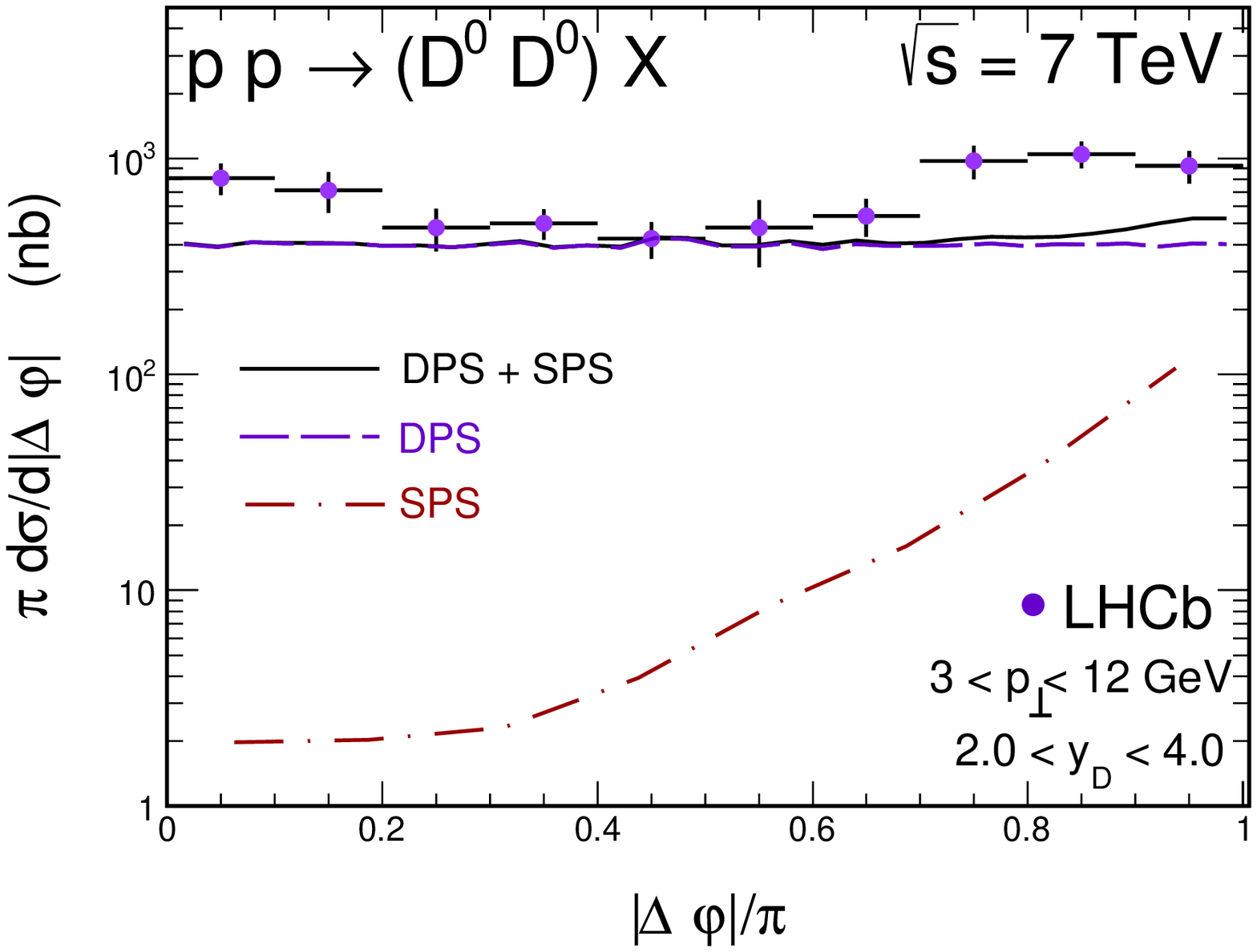}
\end{center}
\vskip -0.3cm
\caption{Rapidity distance between two $D^0$ mesons (left panel) 
and corresponding azimuthal correlations (right panel).
}
\label{fig:DPS_ydiffandphid}
\end{figure}

In real experiments one measures rather $D$ mesons than charm 
quarks/antiquarks. In Fig.~\ref{fig:DPS_ydiffandphid} we show 
resulting distributions in
rapidity distance between two $D^0$ mesons (left panel) and
corresponding distribution in relative azimuthal angle (right
panel). The DPS contribution (dashed line) dominates over the single
parton scattering one (dash-dotted line). The sum of the two
contributions is represented by the solid line. We get a reasonable
agreement with the LHCb experimental data \cite{Aaij:2012dz}.

\begin{figure}
\begin{center}
\includegraphics[width=5.5cm]{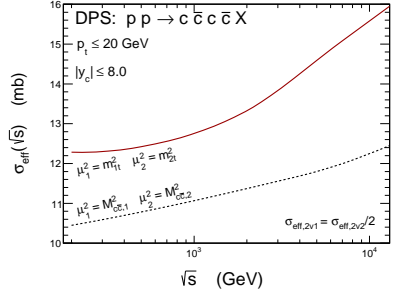}
\end{center}
\vskip -0.3cm
\caption{
Energy and factorization scale dependence of $\sigma_{eff}$
for $c \bar c c \bar c$ production as a consequence of existence
of the two DPS components. 
In this calculation
$\sigma_{eff,2v2}$ = 30 mb and $\sigma_{eff,2v1}$ = 15 mb.
}
\label{fig:sig_eff_charm}
\end{figure}

In the presence of the (single) parton splitting (2v1) contribution 
the situation becoming more complicated \cite{GMS2014}. It was shown 
that the ratio of the 2v1-to-2v2 depends on rapidity of produced
quark/antiquark and collision energy.
In Fig.~\ref{fig:sig_eff_charm} we show the empirical $\sigma_{eff}$,
for double charm production in the case when conventional and single
parton splitting contributions are added together.
The effective parameter $\sigma_{eff}$ rises with the centre-of-mass
energy. A sizeable  difference of results 
for different choices of scales can be observed in addition.
Observation of such an effect would require very precise experimental
data for a few center of mass energies.

\begin{figure}[!h]
\begin{center}
\includegraphics[width=5cm]{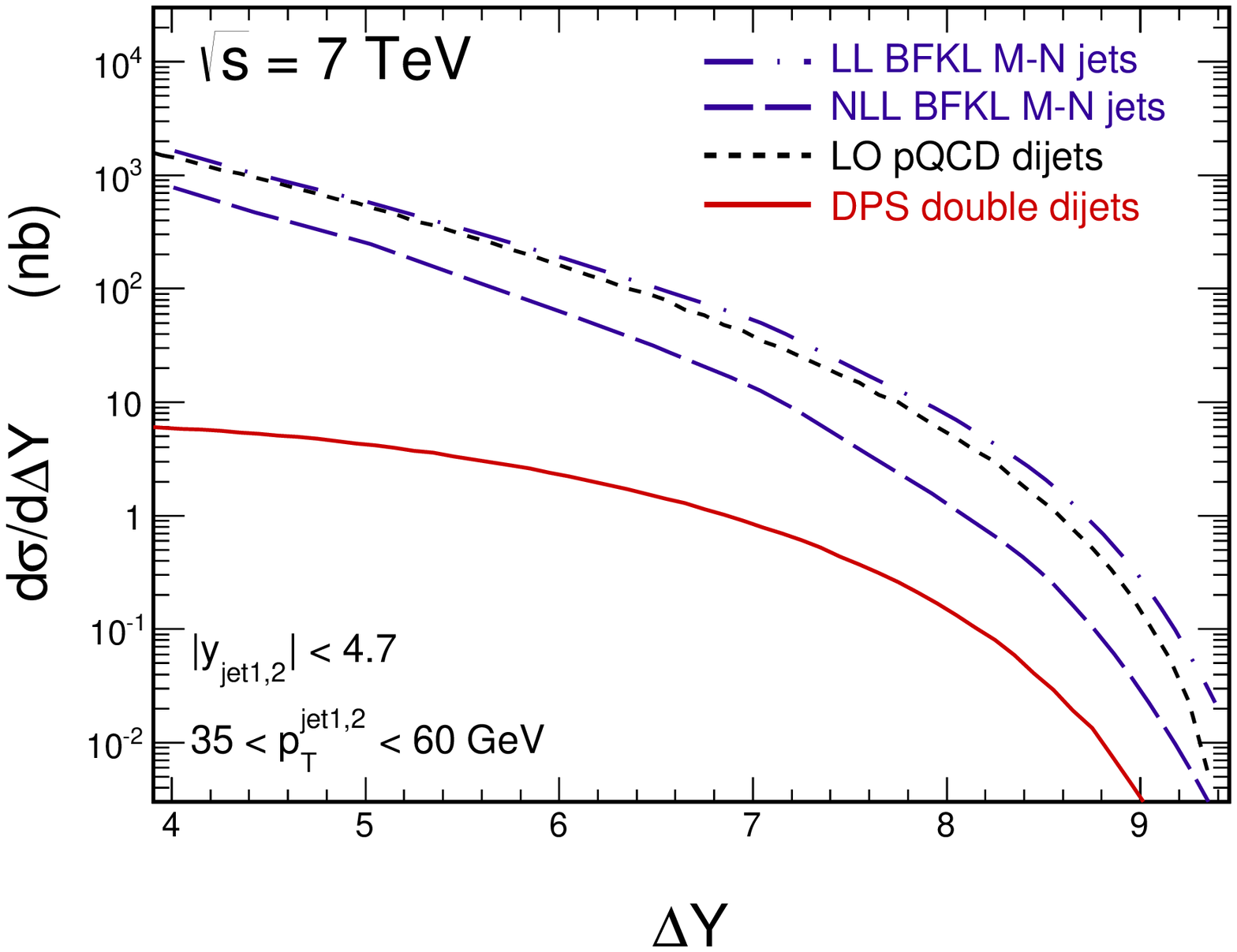}
\includegraphics[width=5cm]{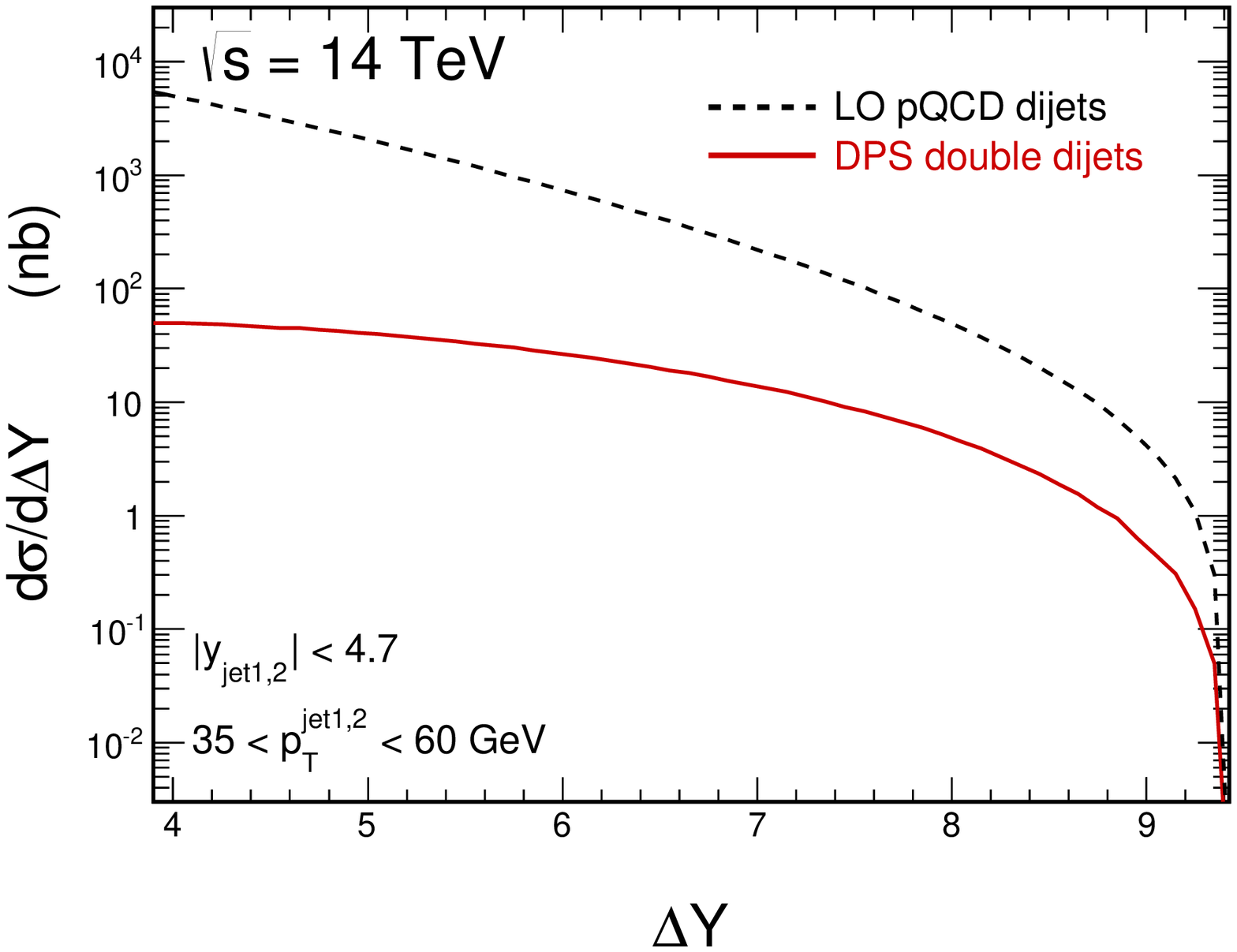}
\end{center}
\vskip -0.3cm
   \caption{
\small Distribution in rapidity distance between jets 
(35 GeV $< p_t <$ 60 GeV).
The collinear pQCD result is shown by the short-dashed line
and the DPS result by the solid line for $\sqrt{s}$ = 7 TeV (left panel)
and $\sqrt{s}$ = 14 TeV (right panel). For comparison we show also
results for the BFKL Mueller-Navelet jets in leading-logarithm 
and next-to-leading-order logarithm approaches from 
Ref.~\cite{Ducloue:2013hia}.
}
 \label{fig:Deltay1}
\end{figure}

Now we proceed to the jets with large rapidity separation.
In Fig.~\ref{fig:Deltay1} we show distribution in the rapidity 
distance between two jets in leading-order collinear calculation
and between the most distant jets in rapidity in the case of four DPS jets.
In this calculation we have included cuts for the
CMS expriment \cite{CMS_private}: $y_1, y_2 \in$ (-4.7,4.7),
$p_{1t}, p_{2t} \in$ (35 GeV, 60 GeV).
For comparison we show also results for the BFKL calculation from
Ref.~\cite{Ducloue:2013hia}. For this kinematics the DPS jets
give sizeable (relative) contribution only at large rapidity distance.
The NLL BFKL cross section (long-dashed line) is smaller than that for 
the LO collinear approach (short-dashed line).

\begin{figure}[!h]
\begin{center}
\includegraphics[width=5cm]{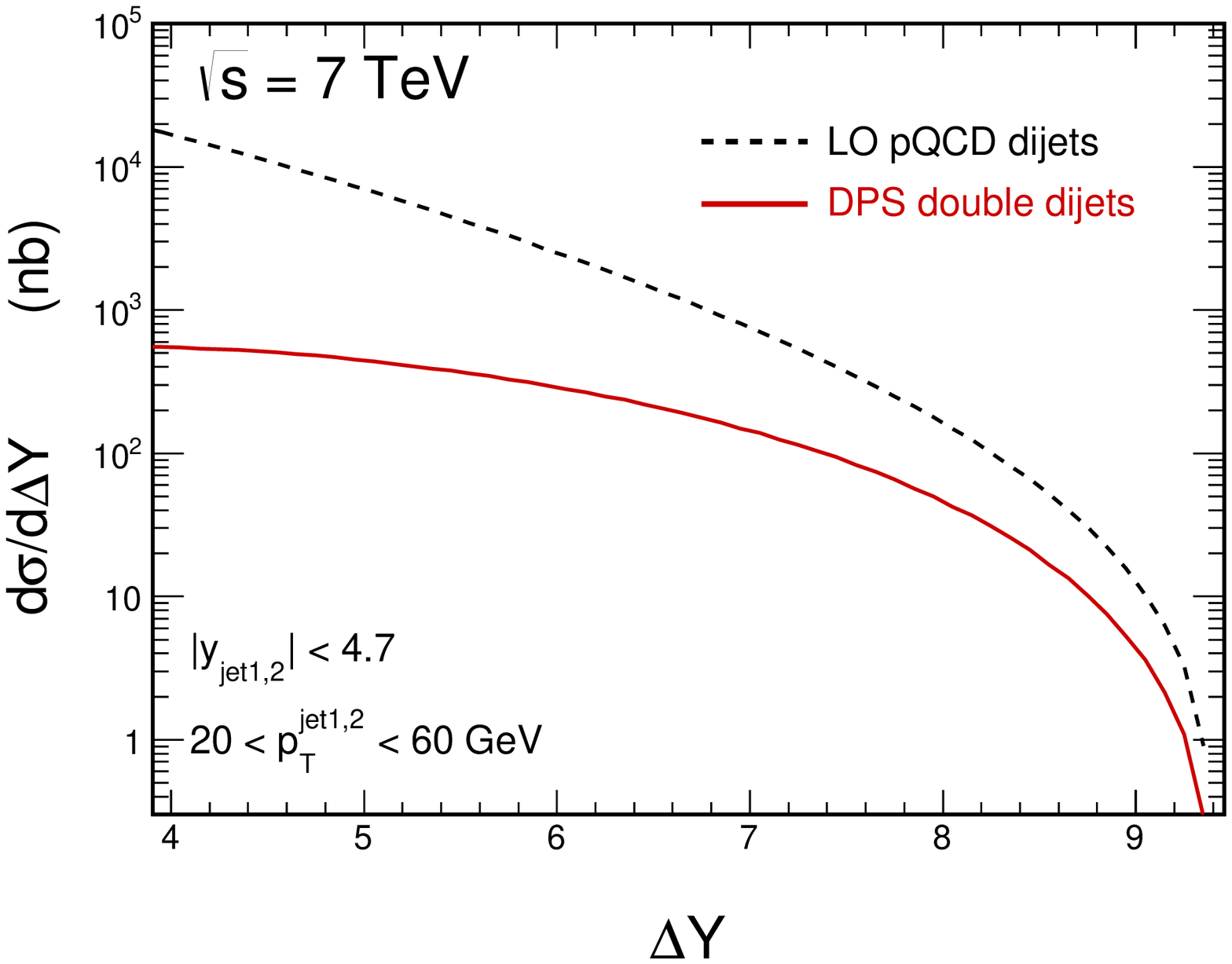}
\includegraphics[width=5cm]{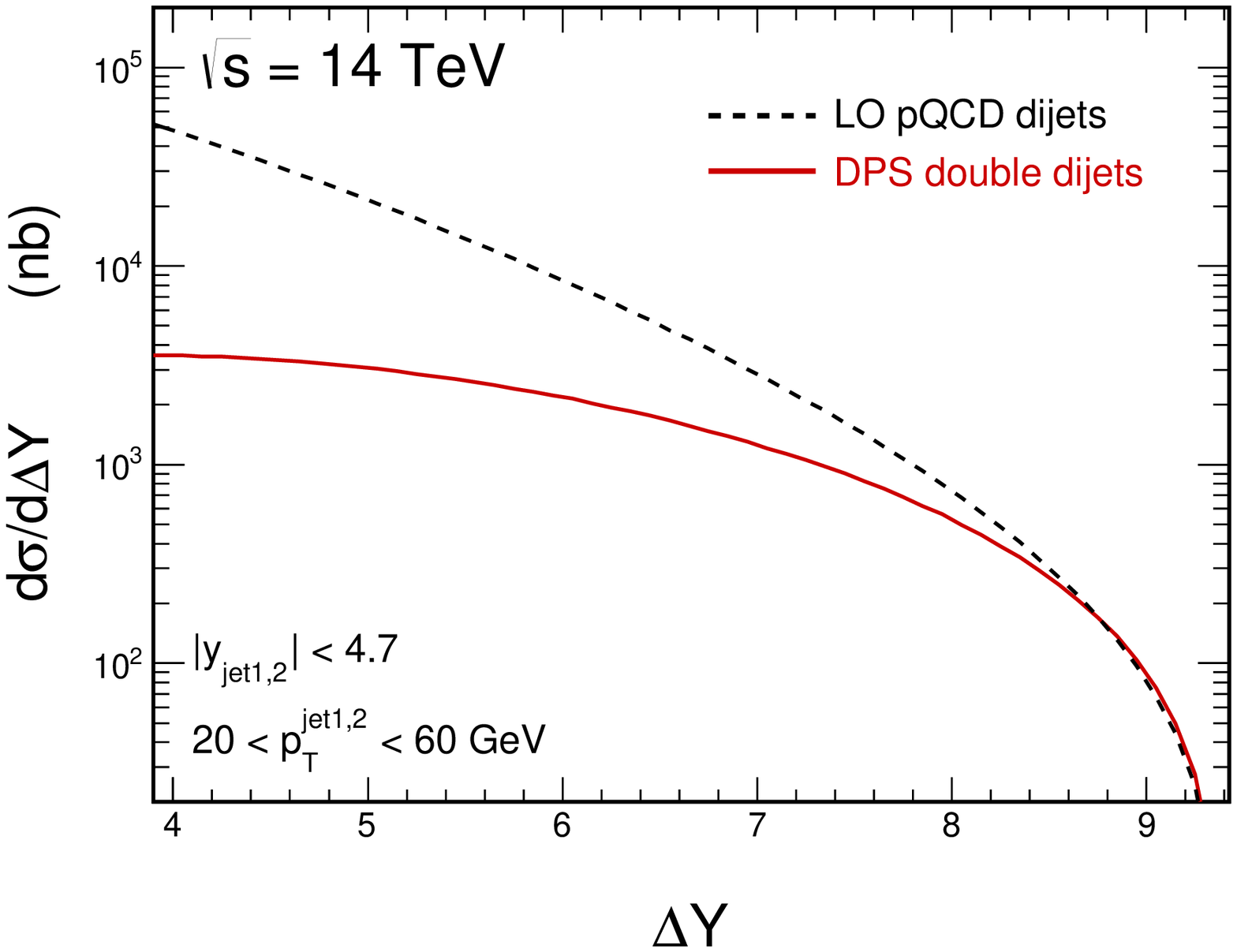} \\
\end{center}
\vskip -0.3cm
 \caption{
\small The same as in the previous figure but now for smaller 
lower cut on minijet transverse momentum.
}
 \label{fig:Deltay-2}
\end{figure}

In Fig.~\ref{fig:Deltay-2} we show rapidity-diffference
distribution for even smaller lowest transverse momenta of 
the "jet". A measurement of such minijets may be, however, difficult. 
Now the DPS contribution may even exceed the standard SPS 
dijet contribution, especially at the nominal LHC energy. 
How to measure such (mini)jets is an open issue. In principle,
one could measure correlations of 
semihard ($p_t \sim$ 10 GeV) neutral pions with the help of 
so-called zero-degree calorimeters (ZDC).

Finally we wish to only briefly mention the situation for production
of two gauge bosons, e.g. $W^+ W^-$ pairs. 
Many mechanisms contribute in this reaction \cite{LSR2015}.
How the situation may look at future high-energy experiments
at the LHC and FCC is illustrated in Table~\ref{table1}.
In this calculation we assumed $\sigma_{eff}$ = 15 mb.
Such an empirical value was extracted from phenomenological studies 
of high-energy processes (also at the LHC) that are, however, 
dominated rather by gluon-gluon processes. 
Whether a corresponding value of $\sigma_{eff}$ for processes 
dominated by quark-antiquark annihilation ($D_{q \bar q'}$ dPDF 
in the cross section formula) is similar is not completely clear at present.
For reference we show the leading-order contribution
due to quark-antiquark annihilation.
Clearly the relative contribution of DPS is growing with collision
energy. The DPS contribution has slightly different characteristics of
the final state than other contributions \cite{LSR2015}.
Detailed studies require to go rather to leptons 
(electrons, muons), study differential distributions and 
consider background contributions. 
This issue was discussed in the literature only briefly (see e.g. 
\cite{Kulesza2010,szczurek_epiphany2015}).

\begin{table}[tb]
\caption{Cross section for $W^+ W^-$ production at different collision 
energies for the dominant $q \bar q$ and DPS contributions.}
\label{table1}
\begin{center}
\begin{tabular}{|l|c|c|}
\hline
       &    $q \bar q$  &     DPS          \\
\hline
8000   &    0.032575    &    0.1775(-03)   \\
14000  &    0.06402     &    0.6367(-03)   \\
100000 &    0.53820     &    0.03832       \\
\hline
\end{tabular}
\end{center}
\end{table}

\section{Conclusions}

We have shown that the cross section for $c \bar c c \bar c$
production grows much faster than the cross section for $c \bar c$ production
making the production of two pairs of $c \bar c$ particularly attractive
in the context of exploring the double-parton scattering processes. 

We have also shown predictions for production of $c \bar c c \bar c$
in the double-parton scattering in the factorized Ansatz with each step
calculated in the $k_t$-factorization approach. 
We have presented also cross sections for the production of 
$D^0 D^0$ (or $\bar D^0 \bar D^0$) pairs of mesons.
The results of the calculation have been compared to recent results of
the LHCb collaboration.
The best agreement with the LHCb data was obtained for 
the Kimber-Martin-Ryskin UGDF. This approach, as discussed already 
in the literature, effectively includes higher-order QCD corrections.
Rather good agreement was obtained for transverse momentum distribution 
of $D^0$ $(\bar D^0)$ mesons and $D^0 D^0$ invariant mass distribution. 
The distribution in azimuthal angle between both $D^0$'s suggests 
that some contributions may be still missing or
the rather crude approximation used in calculating DPS are not sufficient.

We have discussed also so-called parton splitting mechanism in 
the context of the $c \bar c c \bar c$ production.
The sum of the 2v1 and 2v2 contributions behaves
almost exactly like the 2v2 contribution, This leads e.g. to effective
energy dependence of $\sigma_{eff}$.

We have discussed also how the double-parton scattering
effects may contribute to large-rapidity-distance dijet correlations.
Already leading-order calculation provides quite adequate
description of inclusive jet production when confronted with
recent results obtained by the ATLAS and CMS collaborations 
\cite{MS2014_DPSjets}.
We have shown distributions in rapidity distance between
the most-distant jets in rapidity. The relative contribution of 
the DPS mechanism increases with increasing distance in rapidity between
jets.
We have also shown some recent predictions of the Mueller-Navelet jets
in the LL and NLL BFKL framework.
For the CMS configuration our DPS contribution is smaller than 
the dijet SPS contribution and
only slightly smaller than that for the NLL BFKL calculation.
We have demonstrated that the relative effect of DPS can be increased
by lowering the transverse momenta.
A four-jet final states analyses of
distributions in rapidity distance and other kinematical 
observables was performed by us very recently \cite{MS2015}.

Finally we have also mentioned the role of DPS effects in inclusive 
production of $W^+ W^-$ pairs. We have shown that the relative contribution
of DPS grows with collision energy. In experiments one measures
electrons or muons. Whether experimental
identification of the DPS contribution in this case is possible
requires detailed Monte Carlo studies.



\begin{thebibliography}{100}

\bibitem{LP78}
P.V. Landshoff and J.C. Polinghorne, Phys. Rev. {\bf D18} (1978) 3344.

\bibitem{T1979}
F. Takagi, Phys. Rev. Lett. {\bf 18} (1979) 1296.

\bibitem{GSH80}
C. Goebel and D.M. Scott and F. Halzen, Phys. Rev. {\bf D22} (1980) 2789.

\bibitem{H1983}
B. Humpert, Phys. Lett. {\bf B131} (1983) 461.

\bibitem{PT1984}
N. Paver and D. Treleani, Phys. Lett. {\bf B146} (1984) 252.

\bibitem{PT1985}
N. Paver and D. Treleani, Z. Phys. {\bf C28} (1985) 187.

\bibitem{M1985}
M. Mekhfi, Phys. Rev. {\bf D32} 2371;
M. Mekhfi, Phys. Rev. {\bf D32} 2380.

\bibitem{HO1985}
B. Humpert and R. Oderico, Phys. Lett. {\bf B154} (1985) 211.

\bibitem{SZ1987}
T. Sj\"ostrand and M. van Zijl, Phys. Rev. {\bf D36} (1987) 2019.

\bibitem{DH1996}
M. Drees and T. Han, Phys. Rev. Lett. {\bf 77} (1996) 4142.

\bibitem{KS2000}
A. Kulesza and W.J. Stirling, Phys. Lett. {\bf B475} (2000) 168.

\bibitem{FT2002}
A. Del Fabbro and D. Treleani, Phys. Rev. {\bf D66} (2002) 074012.

\bibitem{BJS2009}
E.L. Berger, C.B. Jackson and G. Shaughnessy, Phys. Rev. {\bf D81}
014014 (2010).

\bibitem{GKKS2010}
J.R. Gaunt, C-H. Kom, A. Kulesza and W.J. Stirling, arXiv:1003.3953.

\bibitem{SV2011}
M. Strikman and W. Vogelsang, Phys. Rev. {\bf D83} (2011) 034029.

\bibitem{BDFS2011}
B. Blok, Yu. Dokshitzer, L. Frankfurt and M. Strikman, Phys. Rev. {\bf
  D83} (2011) 071501.

\bibitem{KKS2011}
C.H. Khom, A. Kulesza and W.J. Stirling, Phys. Rev. Lett. {\bf 107}
(2011) 082002.

\bibitem{BSZ2011}
S.R. Baranov, A. M. Snigirev and N.P. Zotov, arXiv:1105.6279.

\bibitem{S2003}
A.M. Snigirev, Phys. Rev. {\bf D68} (2003) 114012.

\bibitem{KS2004}
V.L. Korotkikh and A.M. Snigirev, Phys. Lett. {\bf B594} (2004) 171.

\bibitem{SS2004}
T. Sj\"ostrand and P.Z. Skands, JHEP {\bf 0403} (2004) 053.

\bibitem{GS2010}
J.R. Gaunt and W.J. Stirling, JHEP {\bf 1003} (2010) 005.

\bibitem{GS2011}
J.R. Gaunt and W.J. Stirling, JHEP {\bf 1106} (2011) 048.

\bibitem{DS2011}
M. Diehl and A. Sch\"afer, Phys. Lett. {\bf B698} (2011) 389.

\bibitem{RS2011}
M.G. Ryskin and A.M. Snigirev, Phys. Rev. {\bf D83} (2011) 114047.

\bibitem{DOS2011}
M. Diehl, D. Ostermeier and A. Sch\"afer, JHEP {\bf 1203} (2012) 089. 

\bibitem{MW2012a}
A.V. Manohar and W.J. Waalewijn, Phys. Rev. {\bf D85} (2012) 114009.

\bibitem{MW2012b}
A.V. Manohar and W.J. Waalewijn, Phys. Lett. {\bf B713} (2012) 196.

\bibitem{DK2013}
M. Diehl and T. Kasemets, Phys.Rev. {\bf D91} (2015) 014015.

\bibitem{DKK2014}
M. Diehl, T. Kasemets and S. Kane, JHEP{\bf 1405} (2014) 118.

\bibitem{LMS2012}
 M. {\L}uszczak, R. Maciu{\l}a and A. Szczurek,
 Phys. Rev. {\bf D85} (2012) 094034.

\bibitem{Aaij:2012dz} 
  R.~Aaij {\it et al.}  [LHCb Collaboration],
  J. High Energy Phys. {\bf 06}, 141 (2012);
  J. High Energy Phys {\bf 03}, 108 (2014);
  [arXiv:1205.0975 [hep-ex]].
  
\bibitem{MS2013}
R. Maciu{\l}a and A. Szczurek,
Phys. Rev. {\bf D87} (2013) 074039.  


\bibitem{Mueller:1986ey}
A.~H. Mueller, and H.~Navelet, Nucl.~Phys.~{\bf B282}, 727 (1987).

\bibitem{DelDuca:1993mn}
V.~Del~Duca and C.~R.~Schmidt, Phys.~Rev.~{\bf D49}, 4510 (1994); arXiv:9311290 [hep-ph].

\bibitem{Stirling:1994he}
W.~J.~Stirling, Nucl.~Phys.~{\bf B423}, 56 (1994); arXiv:9401266 [hep-ph].

\bibitem{DelDuca:1994ng}
V.~Del~Duca and C.~R.~Schmidt, Phys.~Rev.~{\bf D51}, 2150 (1995); arXiv:9407359 [hep-ph].

\bibitem{Kim96}
V.~T.~Kim and G.~B.~Pivovarov, Phys.~Rev.~{\bf D 53}, 6 (1996); arXiv:9506381 [hep-ph].

\bibitem{Andersen2001}
J.~Andersen, V.~Del~Duca, S.~Frixione, C.~Schmidt and W.~J~Stirling, J. High Energy Phys. {\bf 02}, 007 (2001); arXiv:0101180 [hep-ph].

\bibitem{Bartels-MNjets}
J.~Bartels, D.~Colferai and G.~Vacca, Eur.~Phys.~J.~{\bf C24}, 83 (2002); Eur.~Phys.~J.~{\bf C 29}, 235 (2003).

\bibitem{Vera:2007kn}
A.~Sabio~Vera and F.~Schwennsen, Nucl.~Phys.~{\bf B776}, 170 (2007); arXiv:0702158 [hep-ph].

\bibitem{Marquet:2007xx}
C.~Marquet and C.~Royon, Phys.~Rev.~{\bf D79}, 034028 (2009); arXiv:0704.3409 [hep-ph].

\bibitem{Colferai:2010wu}
D.~Colferai, F.~Schwennsen, L.~Szymanowski and S.~Wallon, J. High Energy Phys. {\bf 12}, 026 (2010);
arXiv:1002.1365 [hep-ph].

\bibitem{Caporale:2011cc}
F.~Caporale, D.~Y.~Ivanov, B.~Murdaca, A.~Papa and A.~Perri, J. High Energy Phys. {\bf 02}, 101 (2012);
arXiv:1112.3752 [hep-ph].

\bibitem{Ivanov:2012ms}
D.~Y.~Ivanov and A.~Papa, J. High Energy Phys. {\bf 05}, 086 (2012); arXiv:1202.1082 [hep-ph].

\bibitem{Caporale:2012ih}
F.~Caporale, D.~Y.~Ivanov, B.~Murdaca and A.~Papa, Nucl.~Phys.~{\bf B877}, 73 (2013); arXiv:1211.7225 [hep-ph].

\bibitem{Ducloue:2013hia}
B.~Ducloue, L.~Szymanowski and S.~Wallon, J. High Energy Phys. {\bf 05}, 096 (2013);
arXiv:1302.7012 [hep-ph].

\bibitem{Ducloue:2013bva} 
  B.~Ducloué, L.~Szymanowski and S.~Wallon,
  Phys.\ Rev.\ Lett.\  {\bf 112}, 082003 (2014); arXiv:1309.3229 [hep-ph].

\bibitem{DelDuca2014}
V.~Del~Duca, L.~J.~Dixon, C.~Duhr, J.~Pennington, J. High Energy Phys. {\bf 02}, 086 (2014): arXiv:1309.6647.




\bibitem{CMS_private}
I. Pozdnyakov, private communication

\bibitem{MS2014_DPSjets}
R. Maciula and A. Szczurek, Phys. Rev. {\bf D90} (2014) 014022. 

\bibitem{Kulesza2010}
J.R. Gaunt, Ch.-H. Kom, A. Kulesza and W.J. Stirling,
Eur. Phys. J. {\bf C69} (2010) 53,
arXiv:1003.3953 [hep-ph].


\bibitem{LSR2015} 
  M.~Luszczak, A.~Szczurek and C.~Royon,
  J. High Energy Phys. {\bf 02}, 098 (2015);
  [arXiv:1409.1803 [hep-ph]].

\bibitem{GL2014}
K. Golec-Biernat and E. Lewandowska,
Phys. Rev. {\bf D90} (2014) 094032.

\bibitem{KP2013}
W. Krasny and W. P{\l}aczek, Acta Phys. Pol. {\bf B45} (2014) 71,
arXiV:1305.1769 [hep-ph].


\bibitem{CT1999}
G. Calucci and D. Treleani, Phys. Rev. {\bf D60} (1999) 054023.


\bibitem{GMS2014}
J.R. Gaunt, R. Maciula and A. Szczurek,
Phys. Rev. {\bf D90} (2014) 054017.

\bibitem{MS2014} 
  R.~Maciula and A.~Szczurek,
  Phys.\ Rev.\ D {\bf 90}, 014022 (2014)
  [arXiv:1403.2595 [hep-ph]].







\bibitem{Aaij:2011yc} 
  R.~Aaij {\it et al.}  [LHCb Collaboration],
  Phys.\ Lett.\ B {\bf 707}, 52 (2012)
  [arXiv:1109.0963 [hep-ex]].

\bibitem{Aad:2013bjm} 
  G.~Aad {\it et al.}  [ATLAS Collaboration],
  New J.\ Phys.\  {\bf 15}, 033038 (2013)
  [arXiv:1301.6872 [hep-ex]].

\bibitem{SS2012}
 W. Sch\"afer and A. Szczurek,
 Phys. Rev. {\bf D85} (2012) 094029.

\bibitem{Hameren2014} 
  A.~van Hameren, R.~Maciula and A.~Szczurek,
  Phys.\ Rev.\ D {\bf 89}, 094019 (2014); arXiv:1402.6972 [hep-ph].



\bibitem{DL92}
A. Donnachie and P.V. Landshoff, Phys. Lett. {\bf B296} (1992) 227.





\bibitem{HMS2015}
A. van Hameren, R. Maciu{\l}a and A. Szczurek,
arXiv:1504.06490 [hep-ph].

\bibitem{MS2015} 
  R.~Maciula and A.~Szczurek,
  arXiv:1503.08022 [hep-ph].

\bibitem{szczurek_epiphany2015}
arXiv:1504.06491 [hep-ph], a talk at the EPIPHANY2015 conference,
January 2015, Krak\'ow, Poland.

\end{thebibliography}
\end{document}